\documentclass[a4paper,11pt]{article}
\pdfoutput=1 % if your are submitting a pdflatex (i.e. if you have
\usepackage{jcappub} % for details on the use of the package, please
\usepackage[T1]{fontenc} % if needed

\usepackage[utf8]{inputenc}
\usepackage{graphicx}
\usepackage{float}
\usepackage{subfigure}
\usepackage{color}
\def\d{{\mathrm{d}}}
\def\beq{\begin{equation}}\def\eeq{\end{equation}}
\def\bea{\begin{eqnarray}}\def\eea{\end{eqnarray}}

\newfont{\cursive}{pzcmi at 9pt}

\title{Investigating the noise residuals around the gravitational wave event GW150914}

\author[a,b,1]{Alex B. Nielsen,\note{Corresponding author.}}
\author[a,b]{Alexander H. Nitz,}
\author[a,b]{Collin D. Capano}
\author[c]{Duncan A. Brown}

\affiliation[a]{Max-Planck-Institut f\"ur Gravitationsphysik, D-30167 Hannover, Germany}
\affiliation[b]{Leibniz Universit{\"a}t Hannover, D-30167 Hannover, Germany}
\affiliation[c]{Department of Physics, Syracuse University, Syracuse NY 13244, USA}

\emailAdd{alex.nielsen@aei.mpg.de}
\emailAdd{alex.nitz@aei.mpg.de}
\emailAdd{collin.capano@aei.mpg.de}
\emailAdd{dabrown@syr.edu}

\abstract{%
We use the Pearson cross-correlation statistic proposed by Liu and Jackson
\cite{Liu:2016kib}, and employed by Creswell et al. \cite{Creswell:2017rbh}, to
look for statistically significant correlations between the LIGO Hanford and
Livingston detectors at the time of the binary black hole merger GW150914.  We
compute this statistic for the calibrated strain data released by LIGO, using
both the residuals provided by LIGO and using our own subtraction of a
maximum-likelihood waveform that is constructed to model binary black hole
mergers in general relativity. To assign a significance to the values obtained,
we calculate the cross-correlation of both simulated Gaussian noise and data
from the LIGO detectors at times during which no detection of gravitational
waves has been claimed. We find that after subtracting the maximum likelihood
waveform there are no statistically significant correlations between the
residuals of the two detectors at the time of GW150914.}
\begin{document}

\maketitle
\flushbottom

\section{Introduction}
\label{sec:introduction}

The direct detection of gravitational waves from the binary black hole merger
GW150914~\cite{Abbott:2016blz} was a groundbreaking discovery in physics and
astronomy.  An event of this importance warrants broad community scrutiny using
a variety of different techniques.  To that end, the LIGO Scientific and Virgo
Collaborations (LVC) have released a number of data products related to
GW150914 through the Gravitational Wave Open Science Center
(GWOSC)~\cite{Vallisneri:2014vxa, LOSC}.  This includes 4096~seconds of
calibrated strain data~\cite{Abbott:2016jsd,2017PhRvD..96j2001C,Viets:2017yvy}
centered on GW150914, along with the figure data of Ref.~\cite{Abbott:2016blz}.

One of the groups to take on the challenge of independently verifying the LVC
results has raised some concerns about the reliability of the detection
\cite{Creswell:2017rbh}. One feature of the concerns of
Ref.~\cite{Creswell:2017rbh} is the possibility of significant excess
correlations (after the signal is subtracted) between the data streams of the
two LIGO detectors, one of which is in Hanford, WA and the other in Livingston,
LA. Since independence of the background noise between the two LIGO detectors
is a key assumption in assigning statistical significance to GW150914, the
concerns of Ref.~\cite{Creswell:2017rbh} have received considerable attention.
Here, we use the methods of Ref.~\cite{Creswell:2017rbh} to look for
correlations in the LIGO data after subtracting the GW150914 signal. We show
that there are no statistically significant correlations that might call into
question the claim of detection of GW150914.

The analysis of Ref.~\cite{Creswell:2017rbh} is fundamentally different from
the analyses performed by the LVC to produce their scientific results. The LVC
analyses that assign statistical significance to GW150914 are discussed in
Ref.~\cite{Abbott:2016blz} and further detailed in a number of companion
\cite{TheLIGOScientific:2016uux, TheLIGOScientific:2016qqj,
TheLIGOScientific:2016zmo} and methodological papers \cite{Allen:2004gu,
Allen:2005fk, Klimenko:2015ypf, Usman:2015kfa, Canton:2014ena, Messick:2016aqy}
(and references therein). These analyses include both unmodeled and modeled
searches. The unmodeled searches make minimal assumptions about the types of
signals to be searched for \cite{TheLIGOScientific:2016uux, Klimenko:2015ypf}.
The modeled searches use binary black hole template waveforms that are based on
general relativity (GR) to filter the detector data \cite{Allen:2004gu,
Allen:2005fk, TheLIGOScientific:2016qqj, Usman:2015kfa, Canton:2014ena,
Messick:2016aqy}. Final results of both of these techniques rely on careful
monitoring of the detectors' performances to exclude data during known
terrestrial disturbances \cite{TheLIGOScientific:2016qqj}. The modeled searches
bounded the significance of GW150914 to higher values than the unmodeled
searches~\cite{Abbott:2016blz}. This is expected: using templates based on
general relativity helps to exclude unknown instrumental and terrestrial noise
in the detector data. A check for any residual signal after subtraction of a
GR-based template was performed by the LVC in
Ref.~\cite{TheLIGOScientific:2016src}. That analysis found that the
Gaussian-noise hypothesis was preferred over any coherent residual signal
between the two detectors, suggesting that all the measured power is well
represented by the GR prediction for the signal from a binary black-hole
merger.

The independent analysis of Ref.~\cite{Creswell:2017rbh} is based on
calculating the Pearson correlation coefficient \cite{Liu:2016kib}
\label{correlation}
\beq C(\tau;t, \omega) = \int^{t+\omega}_{t} \frac{H(t'+\tau)}{\sigma_{H}}\frac{L(t')}{\sigma_{L}}\d t' \eeq
between the data streams of the two LIGO detectors, $H(t)$ and $L(t)$. Here,
$\sigma_{H,L}$ are the standard deviations of the Hanford and Livingston data,
respectively; $t$ gives the time stamp of the data to analyze, while $\omega$
is a window specifying how much data to use in the correlation. As in
Ref.~\cite{Creswell:2017rbh}, we choose $\omega$ to be $20\,$ms. Some of the results of
Ref.~\cite{Creswell:2017rbh} made use of the source data for Fig. 1 of
Ref.~\cite{Abbott:2016blz}, which is available for download from GWOSC. This
figure consists of three elements. The uppermost row shows the strain data of
the two LIGO detectors around the time of GW150914.  A bandpass filter was
applied to this data to remove high (above $350\,$Hz) and low (below $35\,$Hz)
frequencies, where the detectors are less sensitive. A notch filter was also
applied to remove a number of narrow frequencies. The second row of Fig. 1 of
Ref.~\cite{Abbott:2016blz} shows credible regions for template-based and
unmodeled wavelet-based reconstructions of the signal
along with an example waveform simulated by numerical relativity. The timing,
amplitude, and phase of the numerical waveform was tuned to match the data by
visual inspection for illustrative purposes.  The third row of the figure
shows the result of subtracting the numerical relativity waveform of the second
row from the detector strain data of the first row to produce so-called
residuals, which we will refer to as the ``150914 PRL Residuals''.

\begin{figure}
  \includegraphics[width=\columnwidth]{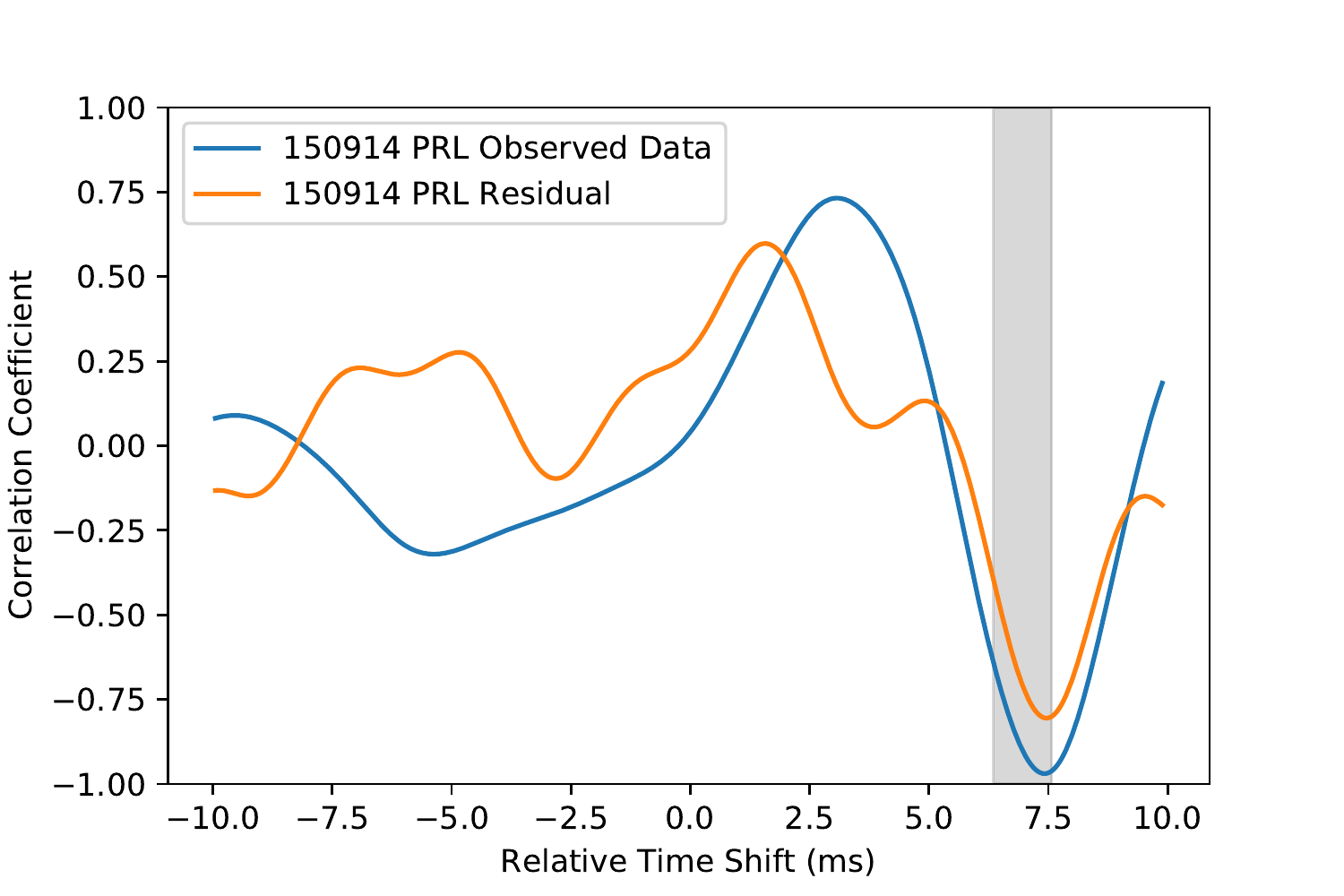}
  \caption{Correlations between timeseries data of the Hanford and Livingston
  detectors that are displayed in Fig.~1 of Ref.~\cite{Abbott:2016blz}. The
  correlations shown are for both the data containing the GW150914 signal (in
  blue) and the example residuals produced after subtracting a numerical
  relativity GR waveform (in orange). As found in Ref.~\cite{Naselsky:2016lay}
  and Ref.~\cite{Creswell:2017rbh}, both the data containing the GW150914
  signal and the example residuals show a significant correlation at a time
  shift corresponding to the time delay between the signal arriving at the
  Livingston detector and then arriving at the Hanford detector. The grey
  shaded region shows a region from 6.5 ms to 7.5 ms that contains the time
  shift between the Hanford and Livingston detectors for the signal observed by
  LIGO.}
  \label{fig:LOSC_correlations}
\end{figure}

It is straightforward to compute the Pearson correlation coefficient using the
150914 PRL Residuals. We use the PyCBC analysis toolkit \cite{pycbc-github} to
reproduce this result here in our Fig.~\ref{fig:LOSC_correlations}; this is
qualitatively similar to the result of Ref.~\cite{Creswell:2017rbh} (their
Fig.~8). A particular concern of Ref.~\cite{Creswell:2017rbh} was that the
correlation coefficient of the 150914 PRL Residuals is peaked at a time-shift
between the Hanford and Livingston data streams that is very close to the
claimed delay time of $6.9^{+0.5}_{-0.4}$ ms for
GW150914~\cite{Abbott:2016blz}. A region encompassing the timing uncertainty is
shown in the plot as a shaded band at $7\pm0.5$ ms in
Fig.~\ref{fig:LOSC_correlations}.

The authors of Ref.~\cite{Creswell:2017rbh} claim that correlations in the residuals of the GW150914 event are concerning enough
to ``raise the possibility that this confirmation [as a gravitational-wave event] may not be completely reliable''. Although
reanalyzing this result is the main topic of this work, here we point out that
a correlation of residuals was not the method employed by the LVC to make the
significance claim for GW150914 in Ref.~\cite{Abbott:2016blz}. The LVC significance
claims for GW150914 are based both on matched-filter searches -- which use GR model
waveforms to match against the data -- and unmodeled searches that look for
excess coherent power between the two detectors. Both of these searches
empirically measure the probability of obtaining chance coincidence between the
detectors by time-shifting the detector data millions of times and reanalyzing.
Neither of these methods assumes that the underlying noise is Gaussian
distributed, as the data is known not to be Gaussian or stationary over long
periods of time \cite{TheLIGOScientific:2016zmo}. Times where one of the
detectors was known to be operating incorrectly were removed from the analysis
ahead of time, as described further in Ref.~\cite{TheLIGOScientific:2016zmo}.
From this background analysis, both the unmodeled and modeled searches obtained
a high statistical significance for GW150914.

Our analysis here differs in several ways from Ref.~\cite{Creswell:2017rbh}.
We use data corresponding to an updated calibration model released by the
LVC in October 2016, labeled as ``v2'' at the GWOSC \cite{LOSC} and ``C02'' by
the LVC, which is slightly different from that plotted in Fig. 1 of
Ref.~\cite{Abbott:2016blz}. We use a different waveform to produce the residual
data. Instead of using the numerical waveform of \cite{Vallisneri:2014vxa}, we
use a maximum likelihood (ML) model waveform from Ref.~\cite{Biwer:2018osg,pycbc-inference-release}. We
estimate the significance of our result using a distribution of uncorrelated
Gaussian noise samples and samples from the real detector noise in time
segments that do not include the GW150914 event. We also investigate the effect
of using whitening rather than bandpass and notching on the correlations.
Whitening of the data more closely follows the methodology used by the LVC to
estimate the significance of GW events. Ref.~\cite{Creswell:2017rbh} also discusses correlations between the Fourier amplitudes and phases around GW150914. We do not investigate that here as it has already been discussed in Ref.~\cite{Green:2017voq}.

\begin{figure}
  \includegraphics[width=\columnwidth]{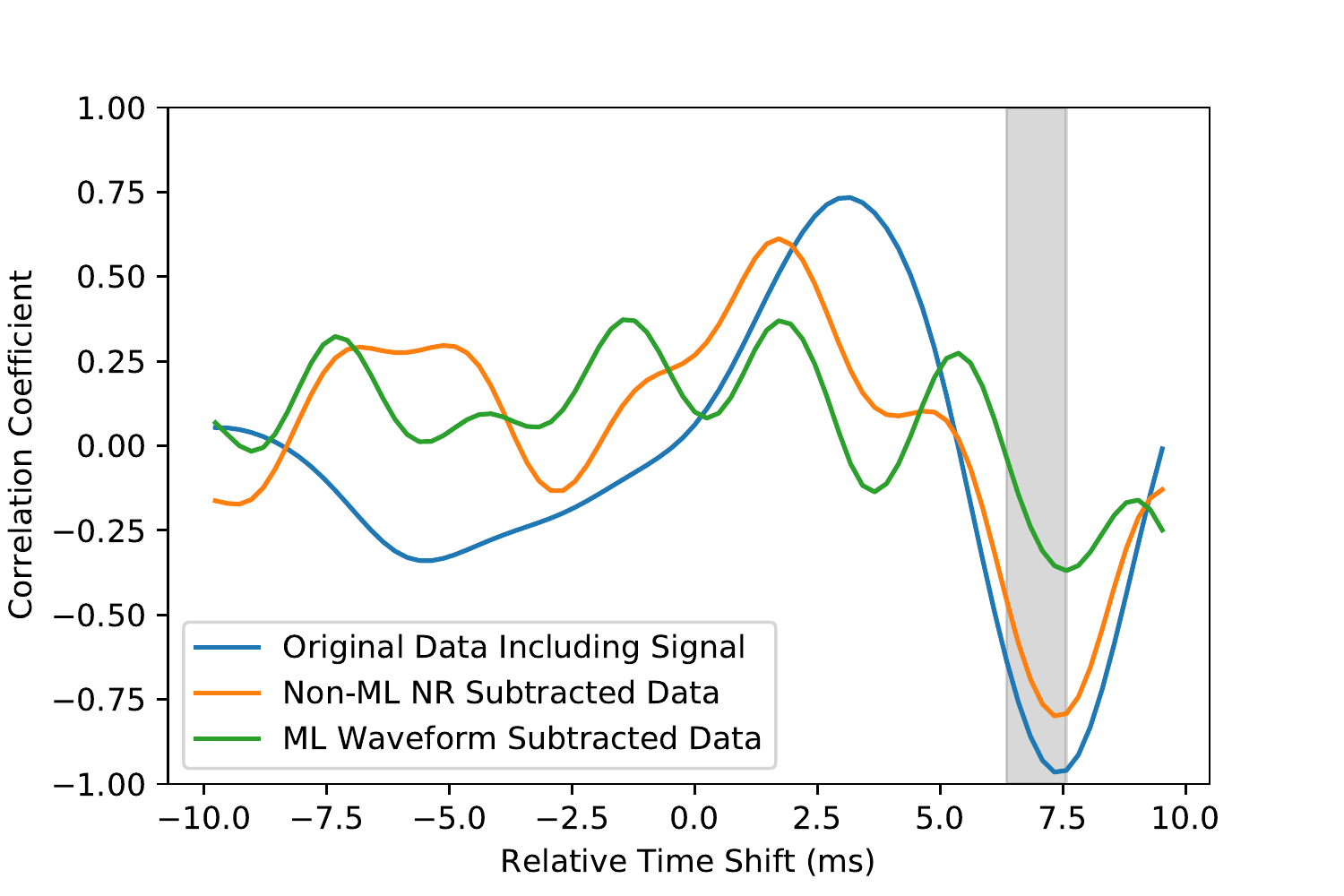}
  \caption{Correlations between the Hanford and Livingston data released at
  Ref.~\cite{LOSC} that has been bandpassed and notched in the same way as used
  to produce Fig.~1 of Ref.~\cite{Abbott:2016blz}. This data has an updated
  calibration compared to data used in Ref.~\cite{Abbott:2016blz}. Therefore
  the orange and blue curves are not exactly identical (although very similar)
  to those plotted in Fig.~\ref{fig:LOSC_correlations}. The figure also
  contains the correlations for residuals constructed by subtracting the
  maximum likelihood model waveform found in
  Ref.~\cite{Biwer:2018osg,pycbc-inference-release}. The correlations for these
  residuals are seen to be significantly lower than for the residuals
  constructed from the example numerical waveform from Fig.~1 of
  Ref.~\cite{Abbott:2016blz}.}
  \label{fig:v2_ML_correlations}
\end{figure}

\section{Residuals after subtracting a maximum-likelihood waveform and their statistical significance}
\label{sec:significance}

The waveform used to produce the residuals of Fig. 1 of
Ref.~\cite{Abbott:2016blz} was a numerical relativity simulation which
approximately fits the data \cite{Lovelace:2016uwp}. Numerical binary
simulations take a long time to run (typically several weeks on a dedicated
computer cluster). They cannot yet be used for a detailed exploration of the
likelihood surface for the parameters involved~\cite{Abbott:2016apu}, such as
the individual masses and spins of the coalescing black holes. Instead, model
waveforms that combine both analytical and numerical relativity information are
used which are considerably faster to generate. These model waveforms have been
shown to agree well with numerical relativity simulations of GW150914
\cite{Abbott:2016apu}.  The LVC used waveform models to estimate the source
parameters of GW150914 \cite{TheLIGOScientific:2016wfe}. Parameters of a
maximum likelihood (ML) model waveform for GW150914 from such a parameter
exploration were released with Ref.~\cite{Biwer:2018osg,pycbc-inference-release}. These values apply to
the IMRPhenomPv2 family of effective precessing spin waveforms
\cite{Hannam:2013oca} that is freely available through the LALSuite package
\cite{LALSuite}. Although the result of Ref.~\cite{Biwer:2018osg} is not
identical to those of the LVC \cite{TheLIGOScientific:2016wfe}
(Ref.~\cite{Biwer:2018osg} did not include a marginalisation over detector
calibration uncertainty), the differences are sufficiently small for our
purposes. The parameters of this waveform lie well within the confidence
intervals given in Ref. \cite{TheLIGOScientific:2016wfe}.  Using this waveform,
suitably projected onto the two detectors for their given location and
orientation, we produce the residuals released as a supplement to this work
\cite{residualrepo}. Due to the difference in data (we use data with an updated
calibration from \cite{LOSC}) and the waveform subtracted (we use the maximum
likelihood waveform from \cite{Biwer:2018osg,pycbc-inference-release}), our residuals are not identical
to the 150914 PRL Residuals.

Although the waveform we subtract was selected by maximising a coherent
Gaussian likelihood function for the two detectors, subsequently subtracting it
from the data will not necessarily produce uncorrelated residuals, since the
coherence algorithm assumes a consistent GR source signal with amplitude and
phasing at the detectors appropriate for a tensorial gravitational wave. It is
possible that there are other correlations between the detectors that are not
coherent or consistent with GR binary mergers. Checking the resulting residuals
with the Pearson correlation coefficient is a consistency test as to whether
such incoherent or non GR-like correlations exist.

We calculate the Pearson correlation coefficient using the same method as
Ref.~\cite{Creswell:2017rbh}. The result is shown in
Fig.~\ref{fig:LOSC_correlations} for the data released for Fig. 1 of
Ref.~\cite{Abbott:2016blz}, and in Fig. \ref{fig:v2_ML_correlations} for the
updated calibration data available from GWOSC~\cite{LOSC}. Both blue curves
depict the correlations with the signal still included. The orange curves show
the correlations after the NR waveform from Ref.~\cite{Abbott:2016blz} is
subtracted. The green curve in Fig. \ref{fig:v2_ML_correlations} shows the
correlations for our residuals after subtracting the maximum-likelihood
waveform of Ref.~\cite{Biwer:2018osg}.

\begin{figure}
  \includegraphics[width=\columnwidth]{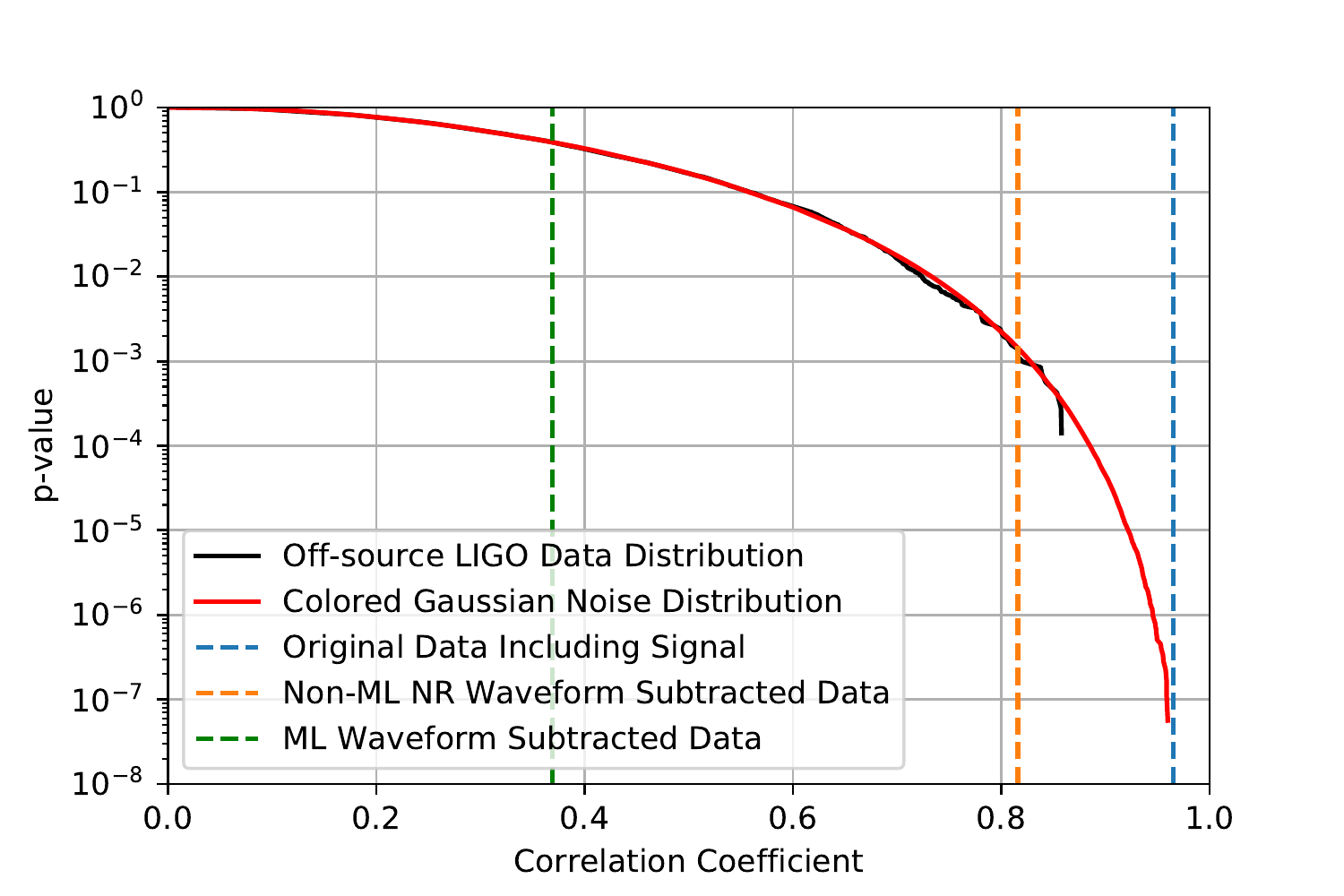}
  \caption{Cumulative distribution of correlations in simulated colored
  Gaussian noise and detector data away from GW150914, using the same bandpass
  and notching, and correlation time window settings as used to produce
  Fig.~\ref{fig:v2_ML_correlations}. The minimum correlation values near the
  time of flight difference in Fig.~\ref{fig:v2_ML_correlations} are plotted as
  vertical lines for reference.}
  \label{fig:bandpassed_background}
\end{figure}

Comparison of the blue and orange curves in Fig.\ref{fig:LOSC_correlations} and
Fig.\ref{fig:v2_ML_correlations} shows that the effect of the updated
calibration on the correlation results is minimal. However, in
Fig.\ref{fig:v2_ML_correlations} the maximum anti-correlation near the detected
time offset of $6.9\,$ms is reduced from 0.799 for the non-ML waveform, to
0.369 for the maximum-likelihood waveform of Ref.~\cite{Biwer:2018osg}. This
result is due to the use of more accurate intrinsic parameters (such as the
masses and spins of the black holes) and extrinsic parameters (that describe
how the source is related to the detectors) of the subtracted template
waveform.

\begin{figure}
  \includegraphics[width=\columnwidth]{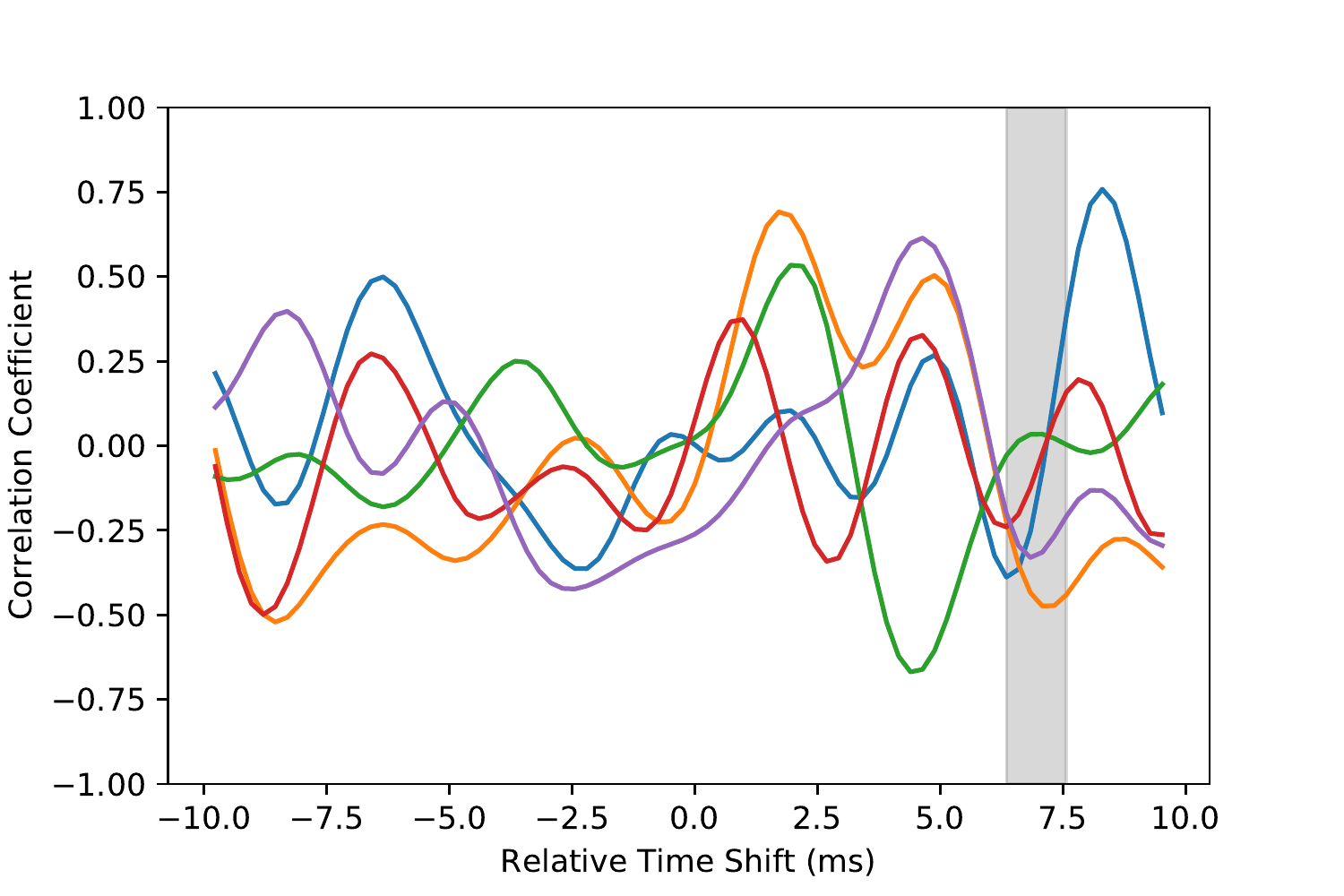}
  \caption{Five examples drawn at random from simulations of uncorrelated
  colored Gaussian noise. The same grey shaded region as
  Fig.~\ref{fig:LOSC_correlations} is shown. This is the same region that is
  used to produce the curves plotted in Fig.~\ref{fig:bandpassed_background}.}
  \label{fig:examples_from_Gaussian_noise}
\end{figure}

It is important to ask how significant any measured correlations are.  To
answer this question we generated simulated colored Gaussian noise, from which
we calculated $O(10^5)$ independent Pearson correlation coefficients. We
performed the same test with $O(10^4)$ samples from the maximum-likelihood
subtracted data, in segments of 61 ms over periods of 256 seconds both before
and after GW150914. For this choice of segments, the $20\,$ms period of data
corresponding to the GW150914 peak correlation is not included.
The Gaussian noise was colored using a PSD estimated from
the real detector data and was then processed in the same way as the data used
in Fig.  \ref{fig:v2_ML_correlations}, with a bandpass and notch filter. The
p-values are calculated by finding the largest correlation or anti-correlation
around a time offset between the detectors of $7 \pm 0.5\,$ms.  (In practice
this is extended to the nearest complete time sample on either side, so that
the actual range runs from 6.35 ms offset to 7.57 ms with sampling at 4096 Hz.)
The results for this background are displayed in Fig.
\ref{fig:bandpassed_background}. Five example realizations, chosen at random
from the distribution, are additionally displayed in
Fig.~\ref{fig:examples_from_Gaussian_noise}.

Figure~\ref{fig:bandpassed_background} shows that the p-value for the
correlation of the non-ML waveform residuals is $\sim0.001$, whereas the
p-value for the ML subtracted residuals is only $\sim0.4$. While the residuals
of Fig.~1 of Ref.\cite{Abbott:2016blz} contain statically significant
correlations at the $0.001$ level, the correlations in the residuals using the
ML waveform of Ref.\cite{Biwer:2018osg} are not statistically significant. The
data with the GW150914 event still included are of course highly significant
compared to the Gaussian background.

It is also clear from Fig.~\ref{fig:bandpassed_background} that the simulated
Gaussian background and the background estimated in real data from times away
from GW150914 do not differ greatly at the level studied here. A
Kolmogorov–Smirnov test on the two distributions has a p-value of 0.91. This
value is consistent with the null hypothesis that they are drawn from the same
distribution. This does not imply that the real data is strictly Gaussian, but
gives a sense of the differences over the time stretch considered (512 seconds
of data centered around GW150914).

\begin{figure}
  \includegraphics[width=\columnwidth]{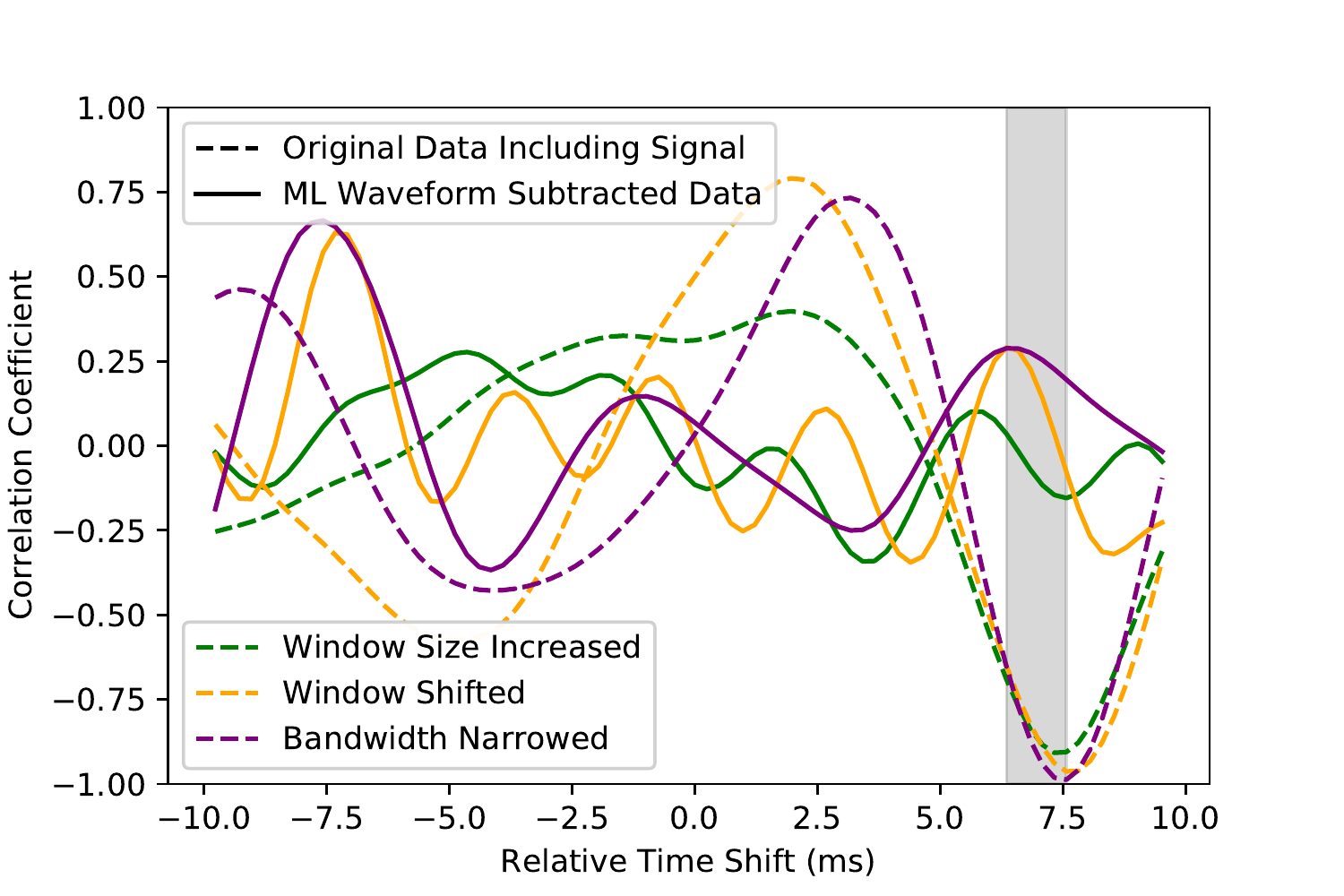}
  \caption{Plot showing the effect of changes to the correlation parameters.
  The dotted lines are data with the signal in, all of which peak strongly at
  the same value of time shift which is largely unaffected by the changes. All
  the solid lines are for the maximum-likelihood waveform subtracted residual
  data and the effect of the changes has a noticeable effect on the value of
  the correlation near the offset time of 7ms. The different colored lines show which
  change was applied to the correlation parameters.}
  \label{fig:changes}
\end{figure}

\section{Robustness of the analysis}
\label{sec:robustness}

Our results in Sec.~\ref{sec:significance} and those of
Ref.~\cite{Creswell:2017rbh} were generated with a particular choice of the
correlation parameters. The frequency bandwidth from $35\,$Hz to $350\,$Hz was chosen to
capture the majority of the signal power for display in Fig.~1 of
Ref.~\cite{Abbott:2016blz}. We used the simple notching and list of spectral
lines taken from the tutorial of Ref.~\cite{LOSC}. The width and start time of
the correlation window corresponds roughly to the four loudest peaks visible in
Fig.~1 of Ref.~\cite{Abbott:2016blz}. None of these choices are set in stone
and other choices are possible. Entirely different methods are used to derive the LVC results
\cite{TheLIGOScientific:2016uux, TheLIGOScientific:2016qqj,
TheLIGOScientific:2016src, TheLIGOScientific:2016wfe}.

In this section, we investigate the effect of modifying some of the correlation
parameter choices.  We separately try tripling the window size, shifting the
start time of the window earlier by $5\,$ms, and narrowing the bandpass filter
to the range $60$--$220\,$Hz. As can be seen in Fig.~\ref{fig:changes}, these
changes do not have a significant effect on the location of the peak
correlation for data containing the GW150914 signal, but they do have a
noticeable effect on the value of the correlation statistic for the
maximum-likelihood waveform subtracted residual data. Indeed, at the low levels
of correlation seen for these residuals, relatively small changes in the
correlation parameters can shift the location of the anti-correlation peak and
even turn an anti-correlation into a correlation near 7ms time separation.
However, none of these correlations in our residual data are at a level that is
statistically significant when compared to simulated Gaussian noise.

\begin{figure}
  \includegraphics[width=\columnwidth]{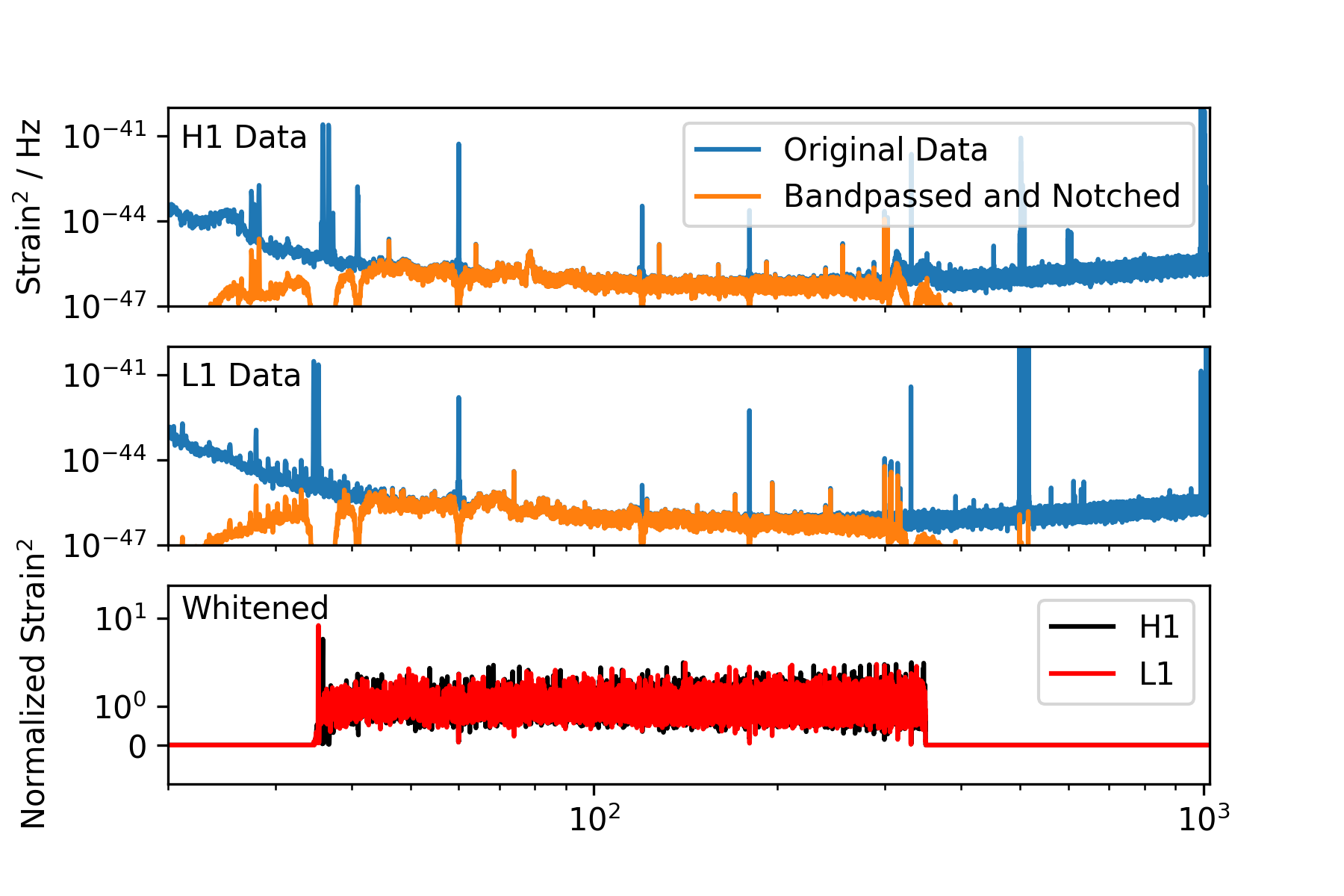}
  \caption{The power spectral density of the data before and after applying the
  band-pass and notching procedure, used to produce Fig.~1 of
  Ref.~\cite{Abbott:2016blz}. It is apparent that not all of the spectral lines
  are removed by the simple choices made to produce Fig.~1 of
  Ref.~\cite{Abbott:2016blz}. The power spectral density after applying the
  whitening procedure employed in this section is also shown for comparison.
  The remaining peak in the power spectral density of the whitened data is due
  to incomplete down-weighting of the calibration lines.}
  \label{fig:bandpass_plus_notch_versus_whiten}
\end{figure}

\begin{figure}
  \includegraphics[width=\columnwidth]{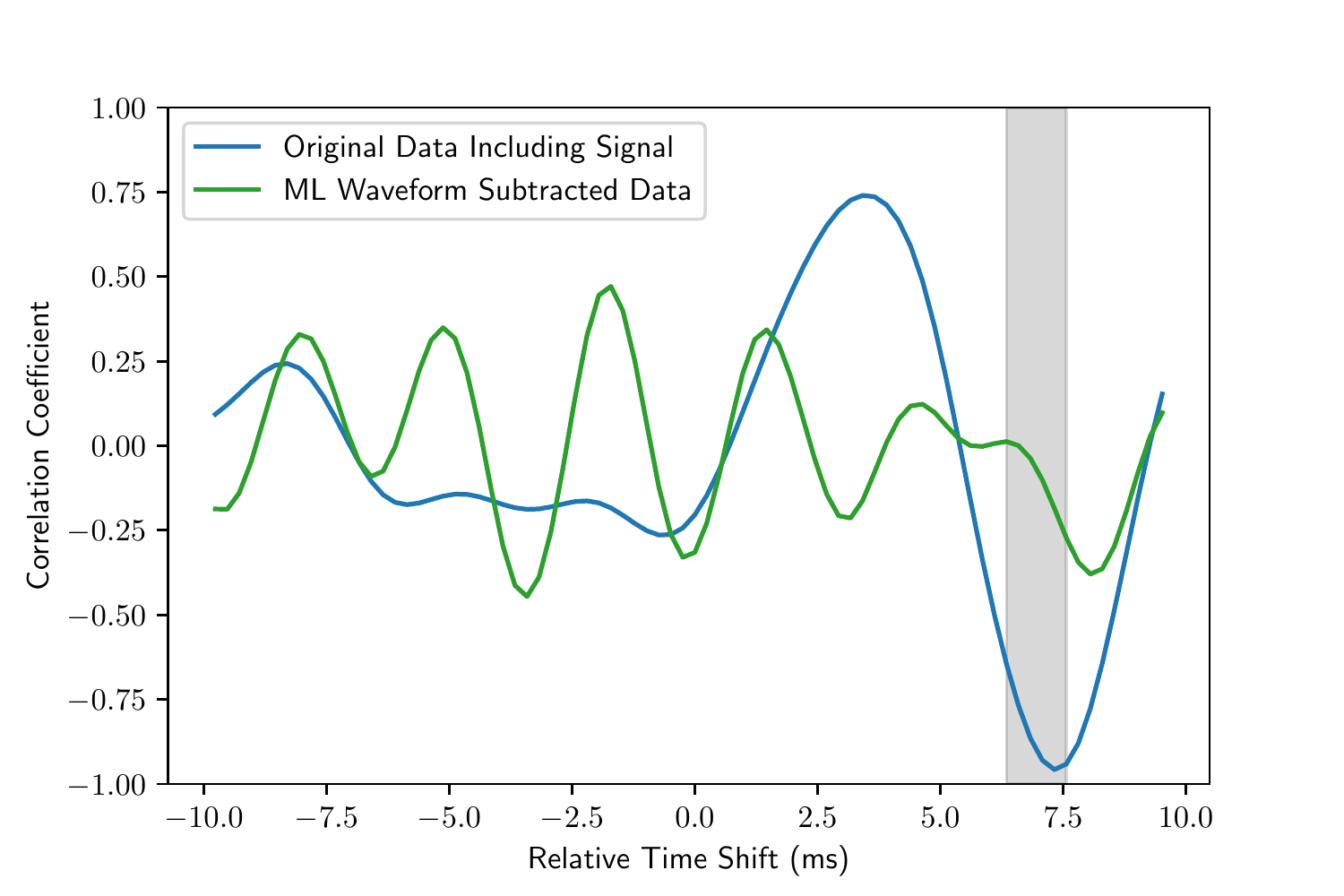}
  \caption{Correlations of residuals after subtracting the
 maximum-likelihood model waveform, whitening, and bandpassing from 35-350Hz. The maximum and minimum peaks in the ML
  residuals (green line) are now away from the time of flight delay of
  $\sim7\,$ms. }
  \label{fig:correlation_whitened}
\end{figure}

\section{Whitening vs bandpassing the data}
\label{sec:whitening}

The LIGO detectors are only sensitive to gravitational waves within a certain
frequency range. At low frequencies, the detectors' sensitivities are dominated
by seismic and thermal noise and at high frequencies by shot noise of the laser
light. The detectors' sensitivities are not white in the frequency
domain~\cite{Martynov:2016fzi}. To make the GW150914 signal visible in Fig.~1
of~\cite{Abbott:2016blz}, a band pass between 35 and $350\,$Hz was applied to the
data.

The LIGO detector data also contain strong spectral lines, small regions of
frequency space where the noise is particularly high. These occur at, for
example, the $60\,$Hz line of electrical power transmission, the violin modes
of the mirror suspensions, and at calibration frequencies used to calibrate the
detectors. In certain cases, these spectral lines will be explicitly correlated
between the Hanford and Livingston detectors for a given offset. To produce
Fig.~1 of Ref.~\cite{Abbott:2016blz}, in addition to bandpassing between
$35\,$Hz and $350\,$Hz, a number of these loud spectral lines were removed by
notching, in particular, the $60\,$Hz line and its harmonics at 120 and
$180\,$Hz.  A list of these notched spectral lines was made available at
Ref.~\cite{LOSC}.  However, to illustrate the signal in Fig.~1 of
Ref.~\cite{Abbott:2016blz}, it was not necessary to remove all spectral lines.
The result of this bandpassing and notching is shown in the upper two rows of
Fig.~\ref{fig:bandpass_plus_notch_versus_whiten}. Although the loudest spectral
lines have been removed, some noticeable lines remain.

The majority of LVC analyses use whitening of the data rather than bandpassing
in producing their results. This includes the pipelines used to assess
significance, both with GR templates~\cite{TheLIGOScientific:2016qqj} and
without~\cite{TheLIGOScientific:2016uux}; to estimate the binary black hole
parameters~\cite{TheLIGOScientific:2016wfe}; and to test the robustness of GR
templates in describing the data~\cite{TheLIGOScientific:2016src}. Whitening
has the advantage that it can suppress spectral lines without a prescribed
list. Figure~\ref{fig:bandpass_plus_notch_versus_whiten} shows a comparison of
the whitened data and the bandpassed and notched data. We see that there remain
several lines in the bandpassed and notched data, and that the power spectrum
varies significantly over the frequency band. Without needing an explicit list,
the whitening procedure has down-weighted the majority of spectral lines.

If instead of bandpassing and notching the detector data -- as done for Fig.~1
of Ref.~\cite{Abbott:2016blz} and used for the residuals analysed by
Ref.~\cite{Creswell:2017rbh} -- the data is whitened before being bandpassed from 35 to 350 Hz and the maximum-likelihood
waveform (also suitably whitened) is subtracted, then we obtain the Pearson
correlation coefficient shown in Fig.~\ref{fig:correlation_whitened}. Similar
to the results of the bandpassing and notching procedure in
Sec.~\ref{sec:significance}, the anti-correlation observed around $6.9\,$ms is
not statistically significant.

\section{Conclusions}
\label{sec:discussion}

We have investigated the question of whether there are statistically
significant correlations between the data of the LIGO Hanford and Livingston
detectors around the time of GW150914. Detailed examinations of potential
sources of detector correlations have also been carried out by the
LVC~\cite{TheLIGOScientific:2017lwt,TheLIGOScientific:2016dpb,
Abbott:2016ezn,TheLIGOScientific:2016zmo}. We have focused on 512 seconds of
data around the time of GW150914 and employed the Pearson correlation
coefficient test to examine whether the results of Ref.~\cite{Creswell:2017rbh}
pose a potential problem with the detection of GW150914. Our residuals are
obtained by subtracting the maximum-likelihood general-relativity based
template found in Ref.~\cite{Biwer:2018osg} from the data. It is worth noting that of this 512 seconds of data, the template has been subtracted from less than 1\% of the time, because the expected signal is negligible outside this range. Both our residuals and maximum-likelihood waveform are available at
\url{www.github.com/gwastro/gw150914_investigation}.

We compute statistical
significance by calculating the Pearson correlation from samples of
uncorrelated colored Gaussian data. We reproduce the result of
Ref.~\cite{Creswell:2017rbh} for LIGO data containing the GW150914 signal, but
find no statistically significant correlations after subtracting the
maximum-likelihood waveform from the data. Although our test statistic and
residual data are different, our findings are consistent with the study of
residual data in Ref.~\cite{TheLIGOScientific:2016src}. Furthermore, a similar
analysis was also performed in Ref.~\cite{Green:2017voq} which likewise concluded
that the residual data was consistent with noise. In addition, Ref.~\cite{Green:2017voq} discussed the concern raised in Ref.~\cite{Creswell:2017rbh} about the Fourier phases. We therefore find no
reason to doubt the significance statements reported in
Refs.~\cite{Abbott:2016blz, TheLIGOScientific:2016pea, Nitz:2018imz} and the
conclusion that GW150914 is a gravitational-wave signal.

\acknowledgments
We thank Sylvia Zhu, Sebastian Khan, Peter Shawhan, Martin Green, and John Moffat for their comments.
We acknowledge the Max Planck Gesellschaft for support.
ABN and DAB thank Andrew Jackson, Hao Liu and Pavel Naselsky for helpful discussions and the 2017
Kavli Summer Program in Astrophysics at the Niels Bohr Institute in Copenhagen and DARK University of
Copenhagen for support during this work. The 2017 Kavli Summer Program program was supported by the the
Kavli Foundation, Danish National Research Foundation (DNRF), the Niels Bohr International Academy and DARK.
DAB thanks Will Farr for helpful discussions and NSF award PHY-1707954 for support.

\appendix
\section{Maximum likelihood IMR waveform parameters}
\label{sec:IMRparams}

We list here the parameters of the general relativity model template that we subtract from the data around GW150914. These values correspond to the maximum likelihood values of \cite{Biwer:2018osg}. They are constructed using the phenomenological inspiral-merger-ringdown waveform family IMRPhenomPv2 \cite{Hannam:2013oca} which is freely available as part of LALSuite \cite{LALSuite}. These results are consistent with the posteriors released at GWOSC~\cite{LOSC} by the LVC as further discussed in \cite{Biwer:2018osg}.
\begin{table}[H]
    \label{tab:table2}
    \begin{tabular}{c|c|c} % <-- Alignments: 1st column left, 2nd middle and 3rd right, with vertical lines in between
      \textbf{Parameter} & \textbf{Description} & \textbf{ML value} \\
      \hline
    $m_1$ & Detector-frame mass of the larger black hole & $39\,\mathrm{M}_\odot$ \\
    $m_2$ & Detector-frame mass of the smaller black hole & $32\,\mathrm{M}_\odot$ \\
    $a_{1}$ & Dimensionless spin magnitude of the larger component & 0.977 \\
    $\theta_1^\mathrm{a}$ & Azimuthal angle of the larger component's spin & 3.6 (rad) \\
    $\theta_1^\mathrm{p}$ & Polar angle of the larger component's spin & 1.6 (rad) \\
    $a_{2}$ & Dimensionless spin magnitude of the smaller component & 0.189 \\
    $\theta_2^\mathrm{a}$ & Azimuthal angle of the smaller component's spin & 3.44 (rad) \\
    $\theta_2^\mathrm{p}$ & Polar angle of the smaller component's spin & 2.49 (rad) \\
    $t_c$ & Geocentric coalescence time (GPS seconds) & 1126259462.4176 \\
    $d_L$ & Luminosity distance & 480 Mpc \\
    $\alpha$ & Right ascension & 1.57 (rad) \\
    $\delta$ & Declination & -1.27 (rad) \\
    $\psi$ & Polarization & 5.99 (rad) \\
    $f_0$ & Starting frequency of the waveform & 10 Hz \\
    $f_{\mathrm{ref}}$ & Reference frequency & 20 Hz \\
    $\iota$ & Inclination of the binary at $f_{\mathrm{ref}}$ & 2.91 (rad) \\
    $\phi_c$ & Reference phase at $f_{\mathrm{ref}}$ & 0.69 \\
    $\Delta \phi$ & Waveform's phase with respect to $\phi_c$ & -0.92 \\
    \end{tabular}
\caption{Parameters of the maximum likelihood waveform. Values are rounded to
arbitrary precision. See the data release associated with this paper
\cite{residualrepo} for exact values used.}
\end{table}

\bibliographystyle{JHEP}

\bibliography{GW150914_correlations}

\end{document}